\documentclass[twocolumn, showpacs, preprintnumbers, amsmath,amssymb]{revtex4}

\usepackage{graphicx}
\begin{document}
\title{Off-critical Casimir effect in Ising slabs
for symmetric\\ boundary conditions in spatial dimension $d=3$}
\author{Z.~Borjan\dag}
\author{P.J.~Upton\ddag}
\affiliation{\dag Faculty~of~Physics,~University~of~Belgrade,~
P.O.Box~368, 11001~Belgrade,~Serbia,}
\affiliation{\ddag Department~of~Mathematics~and~Statistics,
The~Open~University,\\
Walton~Hall, Milton~Keynes, MK7~6AA,~England}

\date{\today}

\begin{abstract}
Extended de Gennes-Fisher (EdGF) local-functional method has been
applied to the thermodynamic Casimir effect {\it away} from
the critical point for systems in the Ising universality class
confined between parallel plane plates with symmetric boundary
conditions (denoted $(ab)=(++)$). Results on the universal scaling
functions of the Casimir force $W_{++}(y)$ ($y$ is a
temperature-dependent scaling variable) and Gibbs adsorption $\tilde
G(y)$ are presented in spatial dimension $d=3$. Also, the mean-field
form of the universal scaling function of the Gibbs adsorption
$\tilde G(y)$ is derived within the local functional theory.
Asymptotic behavior of $W_{++}(y)$ for large values of the
scaling variable $y$ is analyzed in {\it general} dimension $d$.
\end{abstract}


\maketitle

The Casimir effect in quantum or statistical physics refers to long-range forces
that emerge due to a confinement on fluctuations.
In statistical physics
these fluctuations are in the order parameter of a
thermodynamic system {\it at} or {\it near} the critical point,
as predicted in 1978 \cite{FdG}. The Casimir force (CF) depends
on the nature of the confined system as well as the boundary
conditions (b.c.) and the geometrical form of the confinement
\cite{DAE,Krech-1}. Much theoretical work has
 examined various surface universality classes for
  Ising systems and classical fluids either exactly {\it at}
criticality or {\it away} from it \cite{Krech-1}. Symmetry breaking
b.c.\ (defined below) are of particular interest for experiments with
critical binary liquid mixtures. Appreciable agreement pertaining to
these systems has been achieved between theory
\cite{Krech-1,BU-L,Krech-2} and recent experiments on the CF {\it
at} criticality in complete wetting films of binary/fluorocarbon
mixture near liquid vapor coexistence \cite{FYP}, with the mean
value of the universal Casimir amplitude (a measure of CF at the
bulk critical point, defined below) most closely
corresponding to earlier prediction of the local-functional theory
\cite{BU-L}, while at the same time encompassing other
theoretical/simulation estimates \cite{Krech-2}. Exact results on
the full-temperature dependence of the CF are available in spatial
dimensions $d=2$ \cite{Evans-1} and $d \ge 4$ (mean field theory)
\cite{Krech-2,Evans-2} for both cases of symmetry breaking b.c.

Although significant theoretical effort has focused on
the universal scaling functions of the off-critical Casimir
effect, knowledge of them is still somewhat incomplete for spatial
dimension $d=3$, even for the relevant Ising universality class.
Pertinent results in this case  refer to films with periodic b.c.,
studied via Monte Carlo (MC) simulations \cite{DK} or by the
field-theoretic approach for Dirichlet, Neumann and periodic b.c.
\cite{KD}, besides recent MC results that now include
 symmetry breaking b.c. \cite{VD} and are most relevant for the
present content. In this Letter a thermodynamic system
in the Ising universality class is considered in the vicinity of the bulk critical
point. The system is confined between two parallel plane plates of
area $A$ separated by distance $L$. We shall consider only those slabs
where an external symmetry-breaking boundary field has been applied
to both plates, i.e.\ a field $h_1$ (respectively, $h_2$) acting on
the plate at $z=0$ (respectively, $z=L$), and assume that fields $h_1$
and $h_2$ are of the same sign, $h_1 h_2>0$, corresponding to the so
called symmetric b.c.

Thermodynamic CF is defined as a generalized force conjugate to
separation $L$ between the plates $F_{\text{Casimir}}(T;L) :=
-\frac{\partial f^{\times}}{\partial L}$, where $f^{\times}(T;L)$ is the reduced
incremental free enegy defined by $f^{\times}(T;L) := \lim_{A \to \infty}
\frac F{k_B T_c A} - L f_{\mathrm{b}}$ for free energy $F$ with
$f_{\mathrm{b}}$ being the reduced bulk free energy.
It is characterized by the property $F_{\text{Casimir}}(T;L)\to 0$
as $L \to \infty$. According to the finite-size scaling theory,
critical phenomena near the bulk critical temperature $T_c$ and bulk field $h=0$ are
governed by universal scaling functions that depend on the ratio $L
/ \xi$ \cite{Fisher,FB,Barber,PF,Privman}, where $\xi$
is the bulk correlation length with $\xi (t,h=0) \approx
\xi_0^{\pm} \vert t \vert^{-\nu}$, as the reduced temperature $t=
\left (T-T_c) \right/ T_c \to 0^{\pm}$, $\xi_0^{\pm}$
nonuniversal amplitudes and $\nu$ a critical exponent. Then the CF
can be expressed in terms of the universal scaling function
$W_{ab}(\cdot)$ \cite{Krech-1}:
\begin{equation}
F_{\text{Casimir}}(T;L)=L^{-d^\star}W_{ab}(y),~~~y=c_1 t L^{1/\nu},
\label{def:W}
\end{equation}
where $c_1$ is a nonuniversal metric factor. The scaling function
$W_{ab}(y)$, having universal shape \cite{Krech-1}, does depend on
the definition of the correlation length. In order to allow for the
``natural" scaling variable $L / \xi$ ($\xi$ is chosen as
true correlation length) to emerge in the local-functional
expressions of ¢$W_{ab}(y)$ in the asymptotic limits $y \to \pm
\infty$, considered shortly, we choose $c_1=1/(\xi_0^+)^{1/\nu}$.
Exactly at the critical temperature $T_c$, the scaling functions
$W_{ab}(\cdot)$ give the universal Casimir amplitudes
\cite{FdG, Krech-1} $A_{ab}$ via $W_{ab}(0)=(d^\star-1)A_{ab}$,
as already considered within
local-functional theory for symmetric and antisymmetric $(+-)$ ($h_1 h_2<0$) b.c.\ \cite{BU-L}.
 Note that $d^\star=\min(d,d_>)$, where $d$ is a spatial dimension and $d_>$ the upper
critical dimension of the system, the Ising
universality class has $d_>=4$.

The purpose of this Letter is to apply EdGF method  introduced by
Fisher and Upton \cite{FU}, in order to examine the Casimir effect
for systems of the Ising universality class under the symmetric
$(++)$ b.c. over the whole temperature range, in particular, in
$d=3$, important in respect to the experiments where more accurate
theoretical analysis of the above quantities has been missing until now.
As a nonperturbative approach, EdGF theory allows for calculation
directly at a fixed spatial dimension, an advantage over
field-theoretic approach in terms of $\epsilon$ expansion.

The local-functional method \cite{FU} asserts that magnetization
profile $m(z)$ in film geometry is given by minimizing a (local)
interfacial functional ${\cal F}[m]$:
\begin{equation}
{\cal F} [m] := \int_0^L {\cal A} (m, \dot m, t, h) dz +
f_1(m_1;h_1) + f_2(m_2;h_2)
\label{def:functional}
\end{equation}
where $m_1= m(z=0),~ m_2 = m(z=L)$ with $f_i = -h_i m_i - g m_i^2/2$
$(i=1,2)$, the usual surface terms which allow for the presence
of external walls (at $z=0$ and $z=L$), and $\dot m = dm/dz$. The
integrand $\cal A$ is assumed to take the form
 which contains {\it only} bulk quantities \cite{FU}: $
{\cal A}(m, \dot m;t,h) = \{ J(m) {\cal G}[\Lambda (m,t,h) \dot m] +
1\} W(m,t,h)$, where $W(m,t,h) = \Phi (m,t) - \Phi (m_{\mathrm{b}},t) - h
(m-m_{\mathrm{b}})$, and $\Phi (m,t)$ is the bulk Helmholtz free energy
density. Spontaneous magnetization is denoted by $m_{\mathrm{b}}$, where
 $m_{\mathrm{b}}=B(-t)^{\beta}$ for $t<0$ and $m_{\mathrm{b}}=0$ for $t>0$ with
$\beta$ a critical exponent and $B$ a nonuniversal amplitude.
The function ${\cal G}(x)$ is required to satisfy several properties
\cite{BU-L,FU}. As before \cite{FU}, we choose $J(m)=1$ and $\Lambda
(m;t,h) := \xi (m;t)/ \sqrt{2 \chi (m;t) W(m;t,h)}$, where $\xi
(m;t)$ and $\chi (m;t)$ are, respectively, the bulk correlation
length and susceptibility of a homogenous system at $(m,t)$.
Mean-field theory $( d > 4 )$ follows from having
$\Phi (m;t)$ take Landau form with $(\xi^2/2 \chi)
(m;t)$ being {\it constant} in $m$ and $t$. For more general $d>1$,
bulk functions have the following analytic scaling forms \cite{FU}:
\begin{subequations}
\begin{eqnarray}
W(m;t,0)&\approx& \vert m \vert^{\delta + 1}Y_{\pm}(m/m_0(t)),
\label{bulky}
\\
(\xi^2/2 \chi)(m;t)& \approx& \vert m \vert^{-\eta
\nu/\beta}Z_{\pm}(m/m_0(t)), \label{bulkz}
\end{eqnarray}
\end{subequations}
in the simultaneous scaling limits $t \to 0^{\pm}$ and $m \to 0$,
where $m_0(t) := B \vert t \vert^{\beta}$, $\eta$ is the critical
bulk correlation function exponent in standard notation.

Minimization of the functional, Eq.~(\ref{def:functional}), yields
the magnetization profile, $m(z)$, which for $h_1>0$, contains a
minimum at $z=z_{+}$ with magnetization $m_{+}:=m(z_+):=m_0(t)w$.
The scaling variable $y$ and $w$ are related solely in terms of
universal quantities:
\begin{equation}
A_2 \vert y \vert^{\nu} = \int_w^{\infty} \frac{ \sqrt{ \tilde
Z_{\pm}(u)/{\tilde Y_{\pm}(u)}}d u}
 {
u^{1+\nu/{\beta}} \vert \hat {\cal
G}^{-1}[1-(w/u)^{1+\delta}\tilde Y_{\pm}(w)/{\tilde
Y_{\pm}(u)}]\vert }
\label{W1}
\end{equation}
where $\hat {\cal G}(x) := xd{\cal G}/dx-{\cal G}$,
$A_2 :=  R_{\chi}\delta/[Q_2\sqrt{2\delta (\delta+1)}]$
is defined by the standard universal amplitudes
\cite{PHA} $R_{\chi}=C_+ B^{\delta-1} D$, $Q_2=(C_+/C_c)
(\xi_c/\xi_0^+)^{2-\eta}$, $C_+$ a nonuniversal
zero-field susceptibility amplitude above the critical
temperature, $D$, $C_c$ and $\xi_c$ defined along the critical isotherm:
$(T=T_c)$, $h\approx D m^{\delta}$, with $C_c$ being the corresponding susceptibility
amplitude $ \chi \approx C_c \vert h \vert^{-(1-1/ \delta)}$ and $\xi_c$ is
defined from $\xi(m;0) \approx \xi_c
\vert h \vert^{-\nu /\beta \delta}$. The universal functions $\tilde Y(\cdot)$ and
$\tilde Z(\cdot)$ are obtained from normalizing $Y(\cdot)$ and $Z(\cdot)$,
respectively. The local-functional calculation of the CF then follows from
$\partial f^{\times}/\partial L = W(m_+)$, which
yields the universal scaling function $W_{++}(y)$ for $d<4$:
\begin{equation}
W_{++}(y) = -A_1 \vert y \vert^{2-\alpha} w^{1+\delta} \tilde
Y_{\pm}(w), \label{W2}
\end{equation}
with another universal constant $A_1 := R_{\chi}
(R_{\xi}^+)^{d^{\star}} / [(1+ \delta) R_c ]$, defined by other
standard universal amplitudes \cite{PHA}
$R_{\xi}^+ = (\alpha A_+)^{1/d^{\star}} \xi_0^+$
and $R_c = \alpha A_+ C_+ / B^2$, where
$\alpha$ is the specific-heat exponent and $A_+$ a nonuniversal
specific-heat amplitude for $t>0$ and $h=0$.
Eqs.~(\ref{W1}) and (\ref{W2}) determine completely universal
$W_{++}(y)$ within the local-functional approach.

Asymptotic behavior of $W_{++}(y)$ as $y\to\infty$ follows from
Eqs.~(\ref{W1}) and (\ref{W2}) by taking $w\to 0$ from which Eq.~(\ref{W1})
yields $y^{\nu} \approx 2 \ln (B_+/w) + O(w)$, where $B_+$ is some
universal constant. Similarly, $W_{++}(y)$ as $y\to -\infty$ is
obtained from (\ref{W1}) and (\ref{W2}) by taking $w\to 1$ yielding
a similar expression for $|y|^{\nu}$ but in terms of $w-1$. Solving for
$w$ and substituting into Eq.~(\ref{W2}) gives the following
\begin{equation}
W_{++}(y)\approx\left\{ \begin{array}{ll}
                       -W_{+,\infty} y^{2-\alpha}\exp(-y^{\nu}),& \text{as $y\to +\infty$};\\
                       -W_{-,\infty} |y|^{2-\alpha}\exp(-U_{\xi}|y|^{\nu}),& \text{as $y\to -\infty$};
                     \end{array}\right. \label{asymW}
\end{equation}
where $U_{\xi}=\xi_0^+/\xi_0^-$, and $W_{\pm,\infty}$ are new universal amplitudes.
The results summarized by
Eq.~(\ref{asymW}) are {\it general} in that that
they hold in arbitrary spatial dimension $d$.  Previous
results, referring to some special cases, such as exact calculations
on the Ising strip \cite{Evans-1}, and on the Ising chain subject to
two identical surface fields, mean-field analysis based on the
Ginzburg-Landau $\varphi^4$ Hamiltonian \cite{Krech-2}, as well as
mean-field treatments of confined fluids \cite{Evans-2}, confirm the
power-law-exponential behavior of $W_{++}(y)$ shown in
Eq.~(\ref{asymW}).

In obtaining these results one can, to a very good approximation,
set ${\cal G}(x) = x^2$ \cite{BU-L}. This also applies to
all subsequent results pertaining to the symmetric b.c.\ and
greatly simplifies the calculations.

Mean-field form of $W_{++}(y)$  in terms of the Jacobi functions
\cite{Krech-2} follows also within local-functional approach from
Eqs.~(\ref{W1}) and (\ref{W2}), when classical values for critical
exponents are employed along with the scaling functions
$Y_{\pm}(\cdot)$ and $Z_{\pm}(\cdot)$ for $d \ge 4$.

{\it Excess (Gibbs) adsorption $\Gamma (t,h)$.} The Gibbs
adsorption, defined by $\Gamma (t,h) = \int_0^L [m(z;t,h) -
m_{\mathrm{b}}(t,h)] dz$ is an integrated measure of the degree of ordering of
spins or, equivalently, in the language of fluids, the amount of
adsorbed substance on the walls \cite{Maciolek}. From the above
definition and the scaling postulate \cite{Krech-1} $m(z,L,T) \approx
m_0(t) \psi_{++}(x,y)$, $x:=z/L$, valid in the scaling limit $t \to
0$, $L\to \infty$, $z \to \infty$, $L-z \to \infty$, follows:
\begin{subequations}
\begin{eqnarray}
\Gamma(t,0)&=&B(\xi_0^+)^{\beta / \nu } L^{1-\beta / \nu} G(y),
\label{gama1}
\\
G(y)&:=&\vert y \vert^{\beta} \int_0^1 [\psi_{++}(x,y) - \Theta
(-y)] dx,
 \label{gama2}
\end{eqnarray}
\end{subequations}
with $G(y)$ universal and $\Theta(\cdot)$ the Heaviside function.
Asymptotically, $G(y) \sim \vert y \vert^{\beta - \nu}$ as $\vert
y \vert \to \infty$, so that $G(y)$ vanishes for large $y$ and
$d<4,~\beta<\nu$. Since $G(y)$ is not smooth at $y=0$, we perfer to
express results in terms of the universal quantity $\tilde G(y)$, defined for
$d<4$ by
\begin{equation}
\tilde G(y) = G(y) + \vert y \vert^{\beta} \Theta (-y),
\label{gamatilde}
\end{equation}
so that $\int_0^L m dz = B (\xi_0^+)^{\beta / \nu} L^{1-\beta /
\nu} \tilde G(y)$. Local functional theory predicts that
for $d<4$:
\begin{equation}
\tilde G(y)= (1/A_2) y^{\beta - \nu} \int_w^{\infty} \frac{ \sqrt{
\frac{\tilde Z_{\pm}(u)}{\tilde Y_{\pm}(u)}}du}{ u^{\nu / \beta}
\sqrt{1-(\frac{w}{u})^{1+\delta} \frac{\tilde Y_{\pm}(w)}{\tilde
Y_{\pm}(u)}}} \label{gamafinal}
\end{equation}

For $d \ge 4~( \beta = \nu = 1/2)$ scaling forms given by
Eqs.~(\ref{gama1},\ref{gama2}) and (\ref{gamafinal}) fail and one
has to redefine them to encompass a logarithmic correction: in the
limit $L \to \infty$ and $t \to 0$, $\Gamma (L,T) \approx B \xi_0^+
(K_1 \ln L + G(y))+ \Gamma_0(t)$, with $K_1$ as a universal
constant, and $\Gamma_0(t)$ a non-universal additive backround
containing both analytic backround terms and singular corrections.
Mean-field results, that follow from Eq.~(\ref{gamafinal}) when
classical values of critical exponents are used along with the
scaling forms $ \tilde Y_+(y) = 1+2/y^2,~\tilde Y_-(y) =
(1-1/y^2)^2,~ \tilde Z_{\pm}(u)=1$ (one reads them off from
Eqs.~(\ref{bulky}, \ref{bulkz}) for $d \ge 4$) can be expressed in
terms of a complete elliptic integral of the first kind:
\begin{equation}
\tilde G(y) = \left \{ \begin{array}{ll} -\sqrt{2} \ln [ 2 K^2(k)]
,~ y=4 (2 k^2-1) K^2(k), \\ 1/2 \le k^2 \le 1 ; \\
-\sqrt{2} \ln [ {2(1-k^2) K^2(k)}] ,~\\
y=-4(1+k^2)K^2(k),~  0 \le k \le 1 ; \\
-\sqrt{2} \ln [ 2K^2(k) ],~y=-4(1-2k^2) K^{2}(k), \\
0 \le k^2 \le 1/2. \end{array} \label{gamamean-field}\right.
\end{equation}
Mean-field universal scaling function $\tilde G(y)$ is shown by
Fig.~\ref{fig:Gtilde} below, together with the $d=3$
result.

\begin{figure}[h]
\includegraphics[width=80mm]{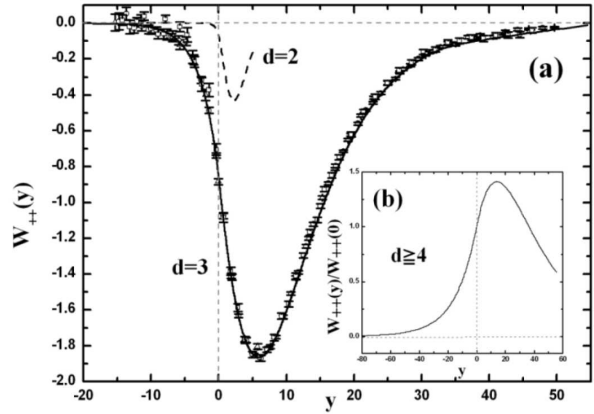} \caption{\label{fig:W++}
Plots of the scaling function $W_{++}(y)$
for the Ising universality class in (a) $d=3$ as obtained by
local-functional theory (solid line) and compared with MC result \cite{VD}
(open circles, rectangles and triangles); exact $d=2$ (dashed line)
\cite{Evans-1}; (b) $d \ge 4$,
presented in a reduced (universal) form
derived for the first time in \cite{Krech-2}. }
\end{figure}

To derive {\it quantitative} predictions at $d=3$ for $W_{++}(y)$
and $\tilde G(y)$ we need to substitute into Eqs.~(\ref{W1}),
(\ref{W2})  and (\ref{gamafinal}) specific values for  bulk critical
exponents along with suitable choices for $Y_{\pm}(y)$ and
$Z_{\pm}(y)$. We represent  bulk scaling functions using parametric
models introduced by Schofield \cite{Scho}. These have been
developed further \cite{FU,Zinn} and are believed to give the best
available fits to bulk data and, by their very construction, to give
scaling functions satisfying  required analyticity properties. For
our purposes, pertaining to the present physical problem situated in
a one-phase region,  the original ``linear" parametric model
\cite{Scho, Fisher} was found to suffice \cite{ZB1}. At $d=3$ we
take $\beta=0.328$ and $\nu=0.632$ (all other exponents follow from
the scaling relations) and a satisfactory fit to the bulk amplitude
ratios, being  properties of  bulk scaling functions, is provided by
taking $b^2=1.30$ and $a_2=0.28$ in the notation of \cite{Zinn}, in
the linear model. The universal scaling function $W_{++}(y)$ that
follows from our calculations in $d=3$ is presented in Fig.~\ref{fig:W++}
together with an earlier exact curve in spatial dimensions
$d=2$ \cite{Evans-1} and $d \ge4$ (mean-field) \cite{Krech-2}, which
not being universal \cite{CD-1} is shown in reduced universal form
$W_{++}(y)/W_{++}(0)$.
 The $d=3$ result confirms
qualitatively similar structure of the CF in $d=3$ with the ones
observed in all the other spatial dimensions. This refers to the
negative sign of the CF for like b.c., smoothness across the whole
interval of scaling variable $y\in\mathbb{R}$
(apart from an $|y|^{2-\alpha}$ singularity at $y=0$, which can be shown to be
quite general)
as expected
based on the fact that the critical point of the film $(T_c(L),
h_c(L))$ is located {\it off} the temperature axis at a non-zero
critical bulk field \cite{Nak}. It also follows from this analysis
that the minimum of $W_{++}(y)$ in $d=3$ is located {\it above} the
critical point as in other dimensions.
 Fig.~\ref{fig:W++} shows that there is striking agreement
between the present calculation of $W_{++}(y)$ and recent MC simulation
results \cite{VD},
with quoted value of $W^{\text {MC}}_{++}(0)=-0.884$ implying that the
value of the Casimir amplitude $A^{\text{MC}}_{++}= -0.442$ is closer
to this and the earlier local-functional result of $-0.42(8)$ \cite{BU-L}
as compared to the previous MC study \cite{Krech-2}.

Numerical predictions for $\tilde G(y)$ in $d=3$, based on the EdGF
Eqs.~(\ref{W1}) and (\ref{gamafinal}) within a parametric
representation, as well as analytic result in the mean-field limit
according to Eq.~(\ref{gamamean-field}), are given by
Fig.~\ref{fig:Gtilde}, showing smooth curves,
 diverging as $\vert y \vert^{\beta}$ for $y \to -\infty$
in accord with the general definition of $\tilde G(y)$.
\begin{figure}[t]
\includegraphics[width=80mm]{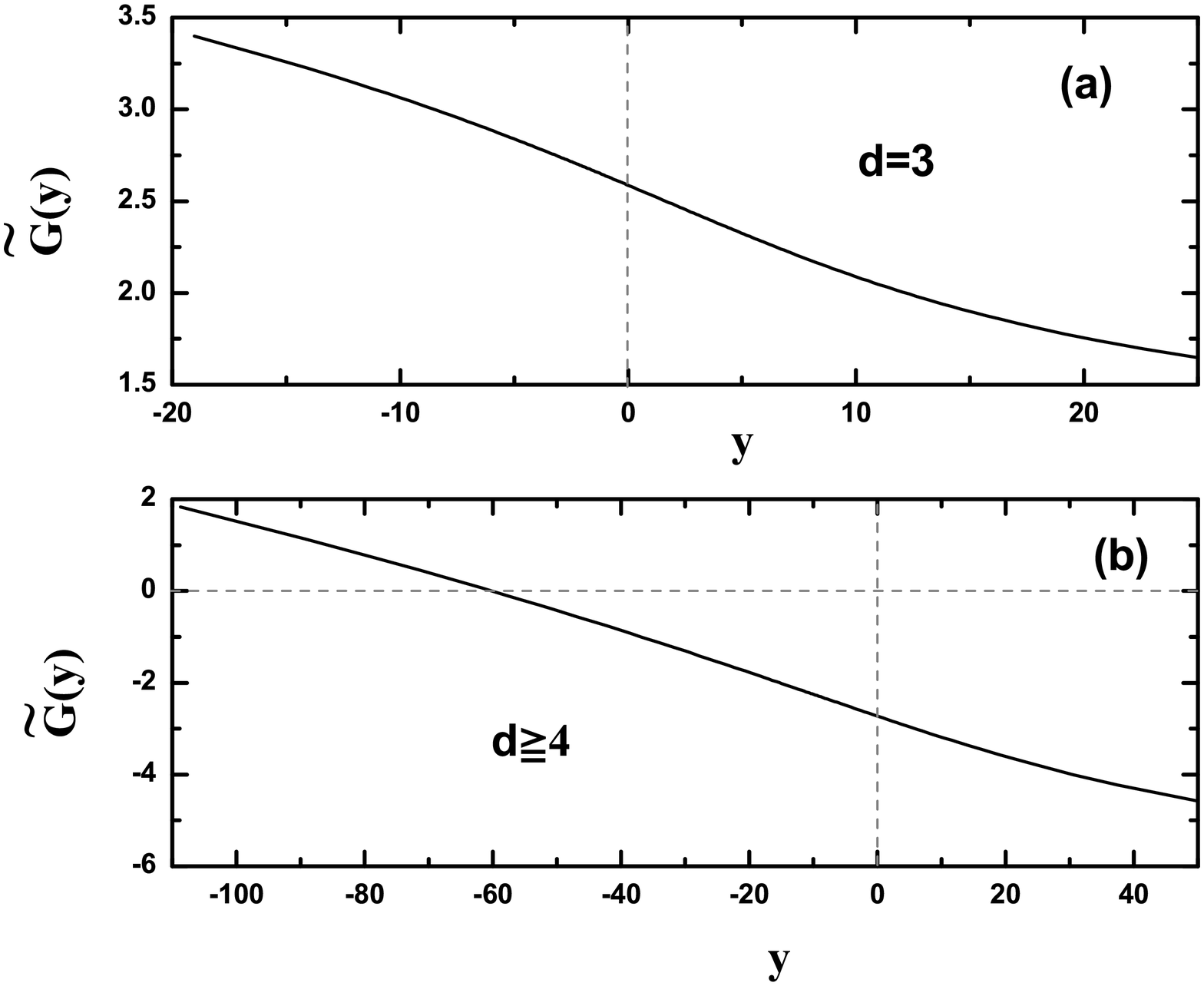} \caption{ \label{fig:Gtilde}Universal scaling function
$\tilde G(y)$ of the excess adsorption $\Gamma$ for the Ising
universality class calculated within local-functional theory (a) in
d=3, by Eqs.~(\ref{W1}) and (\ref{gamafinal}); (b) in $d \ge 4$
according to the EdGF analytic solution (\ref{gamamean-field}). }
\end{figure}

There is also much experimental and theoretical interest in the Casimir
effect for the antisymmetric b.c.\ $(+-)$ with recent MC results
presented for $W_{+-}(y)$ \cite{VD}. In this case, complications
arise in the application of local-functional methods for two main
reasons: (i) the approximation $\mathcal{G}(x)=x^2$ no longer
holds and one needs to use the far more complicated form of
$\mathcal{G}(x)$ as introduced in \cite{BU-L};
(ii) one needs to extend the bulk scaling functions $Y_-(\cdot)$,
$Z_-(\cdot)$ into the two-phase region,
a somewhat ad hoc precedure although possible if one uses
trigonometric parametric models (instead of the linear model)
\cite{FU, Zinn} giving rise to ``nonclassical van der Waals loops''.
However, this more complicated calculation is possible and
forms the subject of ongoing research.

More details will follow in a longer report. We kindly thank
Prof.S.~Dietrich and Dr.O.~Vasilyev for making their  MC data
\cite{VD} available for us that enabled comparisons with the present
result of EdGF theory within Fig.~\ref{fig:W++}.

\bibliography{Off-critical Casimir effect in binary liquid mixtures for symmetric b.c.}

\end{document}